# Superconductivity in the $Nb_2SnC$ compound

A. D. Bortolozo[a], O. H. Sant'Anna[a], M. S. da Luz[a], C. A. M. dos Santos[a], A.S. Pereira[b], K.S. Trentin[c] and A. J. S. Machado[a].

[a] Grupo de Supercondutividade, Departamento de Engenharia de Materiais, Faenquil – Pólo Urbo Industrial Gleba AI 6 s/n, 12600-970, Lorena – SP Brazil.

[b] Instituto de Física e Escola de Engenharia, UFRGS, 91501-970, Porto Alegre, RS, Brazil

[c] Instituto de Física e PGCIMAT, UFRGS, 91501-970, Porto Alegre, RS, Brazil

Abstract

$Nb_2SnC$ is a member of the large family of lamellar materials that crystallize in the hexagonal structure with space group $P6_3/mmc$ which are isomorphs with $Cr_2AlC$, also named H-phase. In spite of the great number of compounds which belong to this family, the superconductivity has been reported only for two cases: $Mo_2GaC$ and $Nb_2SC$. In this work we show that superconductivity can be observed in $Nb_2SnC$ depending on the synthesis method used. The quality of the superconductor is strongly dependent of the synthesis method and the optimal results were reached for samples synthesized at 2.5 GPa and 523 ± 50ºC. This sample showed a critical temperature close to 7.8K, revealed from magnetization and transport measurement, the highest critical temperature reported up to now for an H-phase.

## I - Introduction

About 35 years ago, Nowotny and collaborators published a review paper summarizing the synthesis of a large number of carbides and nitrides which are isomorphs to $Cr_2AlC$.[1] These compounds crystallize in a hexagonal structure, space group $P6_3/mmc$, and follow the stoichiometric ratio $M_2AX$, where M is a transition metal such as V, Ti, Nb, Ta, or Zr, among others; A is an element belonging to the group IV of the periodic table; and X can be carbon or nitrogen. Interesting in this family is its lamellar structure, also called H-phases, which, for the carbides, corresponds to a packing of carbon layers, similar to the graphene layers in graphite, intercalated by layers of the metals M and A obeying the sequence X-M-A-M-X. Alternatively, this structure can be seen as a rocksalt-like MX separated by A group elements such as Al, Ge, Si, S, Sn, etc.[2]

These compounds constitute a new family of layered machinable ceramics and show an unusual combination of properties very attractive to both material scientists and physicists.[1-3] They combine high electrical conductivity (between 0.5 and 14 x $10^6$ $(\Omega.cm)^{-1}$), high thermal conductivity (between 10 and 40 W/mK), Vickers hardness ranging from 2 to 5 GPa, excellent mechanical properties at high temperatures and good machinability. Recent band structure calculations have shown anisotropic electronic properties for some ceramic members of the family such as $V_2AsC$, $Nb_2AsC$, $Nb_2SC$, $Ti_2AlC$, $Ti_2AlN$, and $Ti_3GeC_2$.[4-6] These calculations suggest that the basal plane is responsible for the high electrical conductivity in these compounds. Although this type of electronic behavior is similar to that observed in some well studied superconducting lamellar compounds, there are very few studies in the literature about superconductivity in H-phases. To our knowledge, only $Mo_2GaC$ and $Nb_2SC$ had their superconducting properties investigated.[7-8] In this work, we report a study of the conductivity properties of $Nb_2SnC$ sintered at ambient and high pressure. In the $P6_3/mmc$ structure, the Nb, Sn and C atoms occupy the 4f, 2d and 2a positions, respectively. The layered structure consists of three types of slabs, two Nb-Sn-Nb and one of Nb-C-Nb which obey the sequence C-Nb-Sn-Nb-C. It is also important to emphasize that there are no Nb and Sn atoms placed below or above the C atoms along the c-axis. The results revealed that this compound exhibits superconductivity at ~ 7.8 K. The sample heat treated under high pressure showed sharp superconducting transitions in both transport and magnetization properties.

## II – Experimental Procedure

The samples were prepared using mixtures of graphite, Nb and Sn powders of high purity in the $Nb_2SnC$ stoichiometry. The powder mixture compacted in square form of 10 x 10 $mm^2$ and 2 mm thick, were sealed in a quartz ampoule, and placed in a tubular furnace at 1200$^o$C for 24 hours. After this treatment, the samples were ground and homogenized in an agate mortar, compacted again to the same dimensions mentioned before, and sintered at 1200$^o$C for additional 48 h.

For the high-pressure treatment, a disk shaped sample (6 mm in diameter and 1 mm height) was produced by compaction of the $Nb_2SnC$ powder mixed with a binder (paraffin wax - 2 wt%). This disk was then heated at 270°C for 4 hours to extract the paraffin wax. The high-pressure sintering was carried out in a toroidal-type apparatus.[9] The disk shaped sample was enclosed in a h-BN container and placed inside a graphite sleeve, which acts as an electric heating furnace. After compression to 2.5 GPa, the temperature was raised to 523 ± 50°C and kept at this value for 30 minutes. Then the electric heating system was switched off, and after 10 minutes the pressure was slowly released.

The samples sintered at ambient pressure and under high pressure were characterized by x-ray diffraction in a Seifert diffractometer (model ISO-DEBYEFLEX 1001), using Ni filter $CuK_{\alpha}$ radiation. The samples microstructure was studied by Scanning Electron Microscopy (SEM). Back-scattered electron images (BSE) and x-ray spectroscopic analysis by Energy Dispersive Spectrometry (EDS) and Wavelength Dispersive Spectrometry (WDS) were used to characterize the chemical composition of the sample.

The resistance as a function of the temperature was carried out by the conventional four-point probe method in the temperature interval between 2.0 K and 220 K in an equipment of the Oxford Instruments (MagLab EXA - 9T). Magnetization measurements were performed in a 5 T SQUID magnetometer of the Quantum Design.

## III - Results and Discussion

Figure 1 shows the diffraction pattern of the sample sintered at ambient pressure. We observe the peaks referent of a hexagonal phase pertinent to the space group $P6_3/mmc$ with lattice parameters $a$ = 3.22 Å and $c$ = 13.707 Å.  A single peak, identified by an X in figure 1, was not conclusively identified.

The resistivity behavior for the sample prepared at ambient pressure is shown in the figure 2(*a*). It is possible to note an onset superconducting transition temperature ($T_c$) close to 7.8 K with $\Delta T_c \sim 4.0$ K. These results suggest a small superconducting volume fraction, which is in agreement with the x-ray diffraction pattern (figure 1) that revealed that the sample is not single phase. Also, it is important to emphasize that this sample had a poor densification, probably due to the low temperature used (1200°C) for the ambient pressure sintering. In fact, the magnetization measurements in the ZFC and FC regimes show small magnetic moment intensity as displayed in the figure 2(*b*). In spite of the small superconducting fraction, the onset $T_c$ remains at ~ 7.8 K which is consistent with the transport measurements.

In order to promote the complete reaction and densification, the samples treated in the before conditions, were submitted to a heat treatment at high pressure (2.5 GPa) as discussed in the experimental procedure. This method provided a sample with high densification and with metallic aspect. The x-ray diffraction pattern of this sample (figure 3) corresponds to a single-phase material. The calculated lattice parameters (*a* = 3.245 Å and *c* = 13.77 Å) are in excellent agreement with previous data in the literature for the $P6_3/mmc$ hexagonal phase of $Nb_2SnC$.[3] These results suggest that the high pressure treatment promotes the formation of a single phase sample. This improvement is revealed by the resistive behavior displayed in the figure 4(*a*), which shows a sharp superconducting transition with width of ~ 0.6 K. The magnetic field dependence of the $T_C$, shown in the inset of the figure 4(*a*), allowed us to estimate the upper critical field at zero temperature as ~ 7.8 T. This corresponds to a coherence length ξ(0) ~ 65 Å, if anisotropy is not taken into account.

The magnetization measurement of the sample treated under high pressure is shown in the figure 4(*b*), where a remarkable difference in the FC curves between this sample and the ambient pressure sintered sample can be observed. The onset $T_C$ can be clearly defined with a strong diamagnetism below 7.8 K. Furthermore, the onset critical temperature is consistent with the transport measurement. These results suggest a increasing of the superconducting volume fraction in the sample heat treated under high pressure which was estimated from the FC curve as ~ 32(vol.)% indicating bulk superconductivity in the $Nb_2SnC$ compound. This happened in spite of the high pressure processing to be performed in a temperature significantly lower then that used in the ambient pressure sintering. Although the sintering temperature was lower for high pressure treatment (2.5 GPa), the results shown that this technique improve the diffusion process and densification of the material.

## IV - Conclusions

The compound $Nb_2SnC$ with layered structure exhibits zero resistivity and diamagnetism below 7.8K with bulk type-II superconductivity. The superconducting behavior is extremely dependent on the synthesis method. These characteristics are due to the synthesis and densification in this family of materials. In spite of the lower value of temperature used during the high pressure sintering, the material reaches greater densification and single phase formation promoting the superconducting properties. Furthermore, the results revealed the highest critical temperature reported up to now for an H-phase. Jointly with the $Mo_2GaC$ and $Nb_2SC$ compounds reported in the literature,[7-8] the results shown here also sustained the idea of a new superconducting class of materials that crystallize in a H-phase (prototype $Cr_2AlC$).

**Acknowledgements**

The authors are grateful to FAPESP the financial support through grant number 04/13267-6 and 2005/01257-9

**References**

[1] H. Nowotny, *Prog. Solid. State Chem*., H. Reiss, Ed., 2 (1970) 27.

[2] M. W. Barsoum, T. El-Raghy and M. Radovic, *Interceram*, 49 (2000) 226.

[3] W. Jeitschko, H. Nowotny, F. Benesovsky, Monatshefte Fuer Chemie, 95, 431-435 (1964).

[4] Yanchun Zhou and Zhimei Sun, *Journal of Applied Physics*, 86 (1999) 1430.

[5] Yanchun Zhou and Zhimei Sun, *Physical Review B*, 61 (2000) 12570.

[6] E. Z. Kurmaev, N. A. Skorikov, A. V. Galakhov, P. F. Karimov, V. R. GAlakhov, V. A. Trofimova, Yu. M. Ysrmoshenko, A. Mowes, S. G. Chiuzbaian, M. Neumann, K. Sakamaki, Physical Review B, 71, 014528 (2005).

[7] L. E. Toth, W. Jeitschico and M. Yen, *J. Less-Common Metals*, 10 (1967) 29.

[8] K. Sakamaki, W. Wada, H. Nozaki, Y. Onuki and M. Kawai, *Solid State Communications*, 112 (1999) 323.

[9] M. Eremets, High pressure experimental methods. Oxford University, New York, 1996.

Figure Captions

Figure 1 – Diffractogram of the sample prepared at ambient pressure, which shows peaks of the $Nb_2SnC$ phase and two peaks not identified.

Figure 2 – a) Resistance as a function of temperature revealing superconducting transition at 7.8 K, b) Magnetization in the FC and ZFC regimes displays transition essentially in the same temperature (7.8K).

Figure 3 – Diffractogram of the sample prepared at high pressure revealing the single phase material.

Figure 4 – a) Resistive behavior of the sample prepared at high pressure, showing a sharp superconducting transition close to 7.8K, which reveals the high quality of the sample. The inset of this figure displays magnetoresistance measurements under applied magnetic field $0 \leq \mu_o H \leq 2.0T$ interval. b) *d.c* magnetization as a function of temperature in FC regime which displays magnetic moment signal higher than with the ambient pressure synthesis.

Figure 1 - A. D. Bortolozo et al.

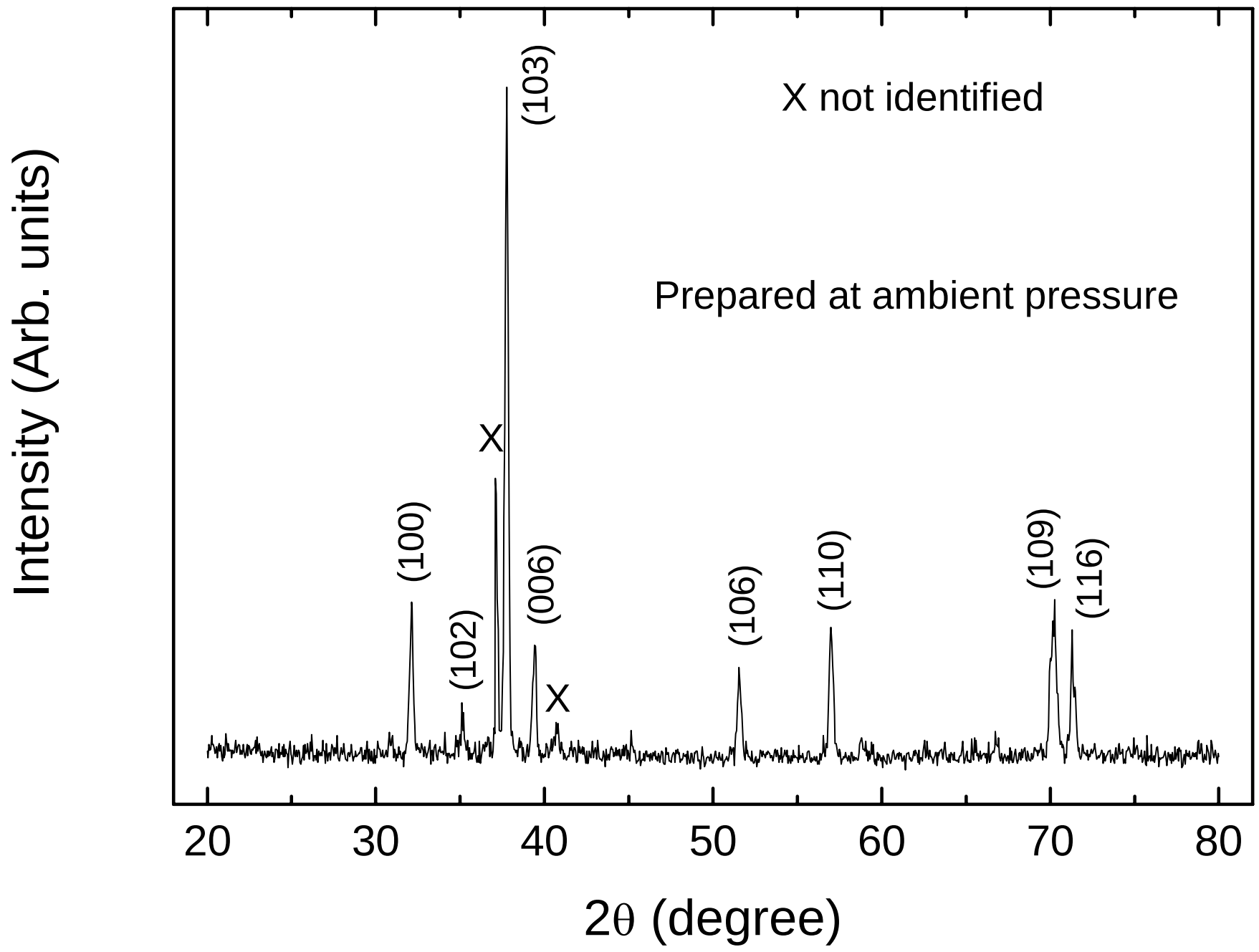

Figure 2 - A. D. Bortolozo

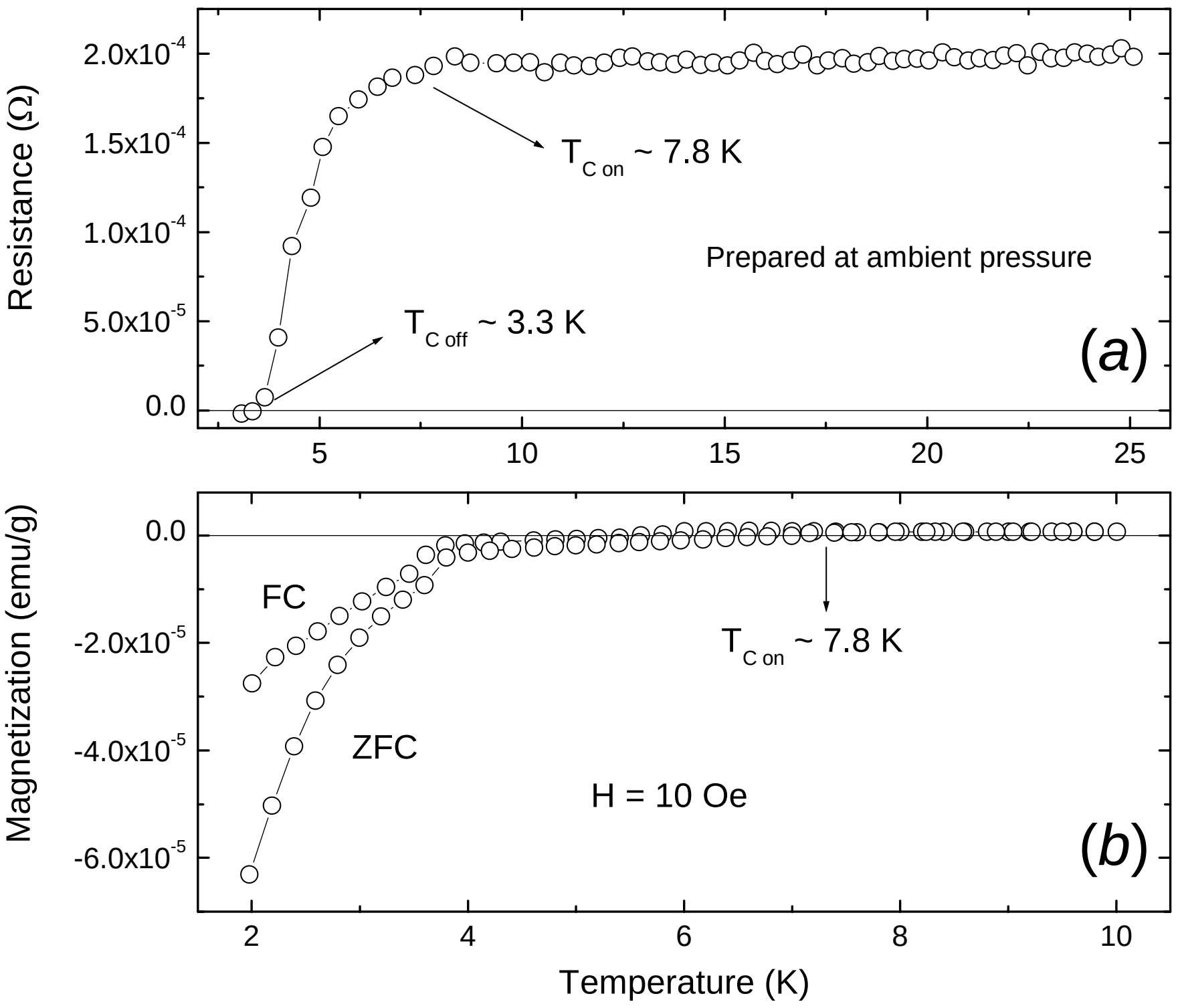

Figure 3 - A. D. Bortolozo et al.

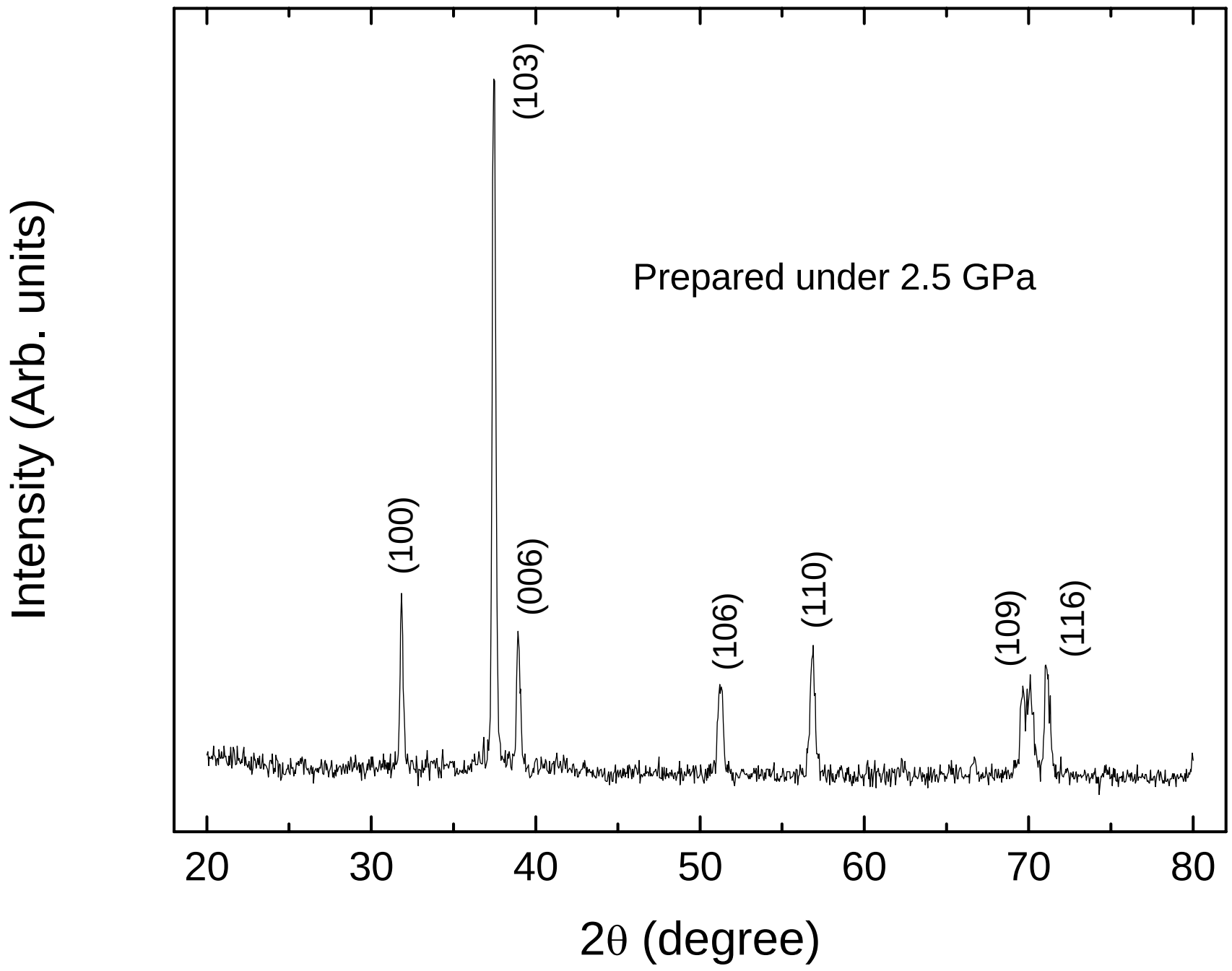

Figure 4 - A. D. Bortolozo et al.

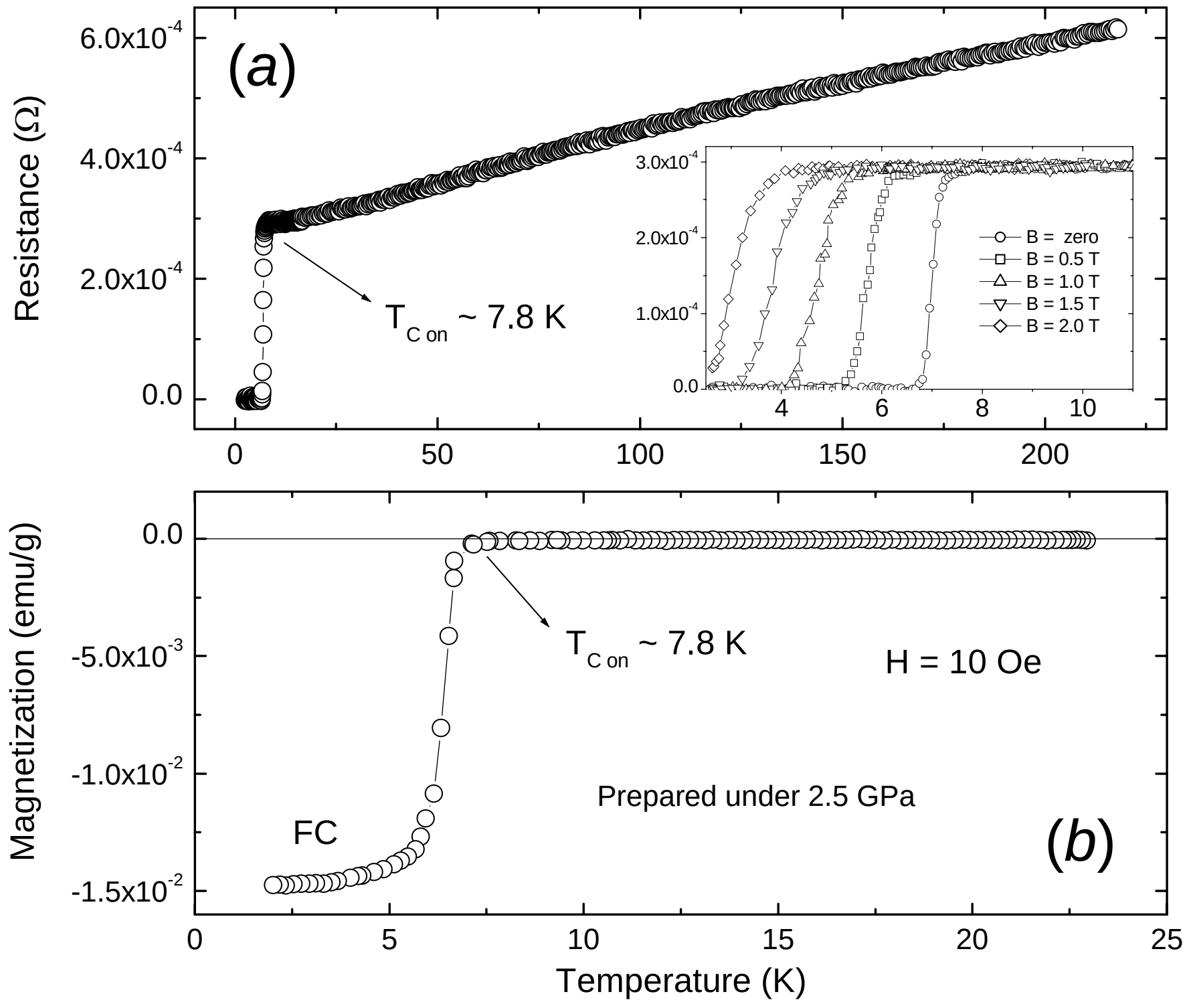